\documentclass{article}
\usepackage{spconf,amsmath,graphicx}
\usepackage{caption, tabularx}
\usepackage{booktabs}
\usepackage{float}
\usepackage{array}  
\usepackage{booktabs} 

\title{SOUND EVENT DETECTION WITH SEQUENTIALLY LABELLED DATA BASED ON CONNECTIONIST TEMPORAL CLASSIFICATION AND UNSUPERVISED CLUSTERING}
\name{Yuanbo Hou$^1$, Qiuqiang Kong$^2$, Shengchen Li$^1$ and Mark D. Plumbley$^2$}
\address{$^1$~Beijing University of Posts and Telecommunications, Beijing, P. R. China \\
	$^2$~Centre for Vision, Speech and Signal Processing, University of Surrey, UK \\
	\{hyb, shengchen.li\}@bupt.edu.cn, \{q.kong, m.plumbley\}@surrey.ac.uk}
\begin{document}
%
\maketitle
%

\begin{abstract}
Sound event detection (SED) methods typically rely on either strongly labelled data or weakly labelled data. As an alternative, sequentially labelled data (SLD) was proposed. In SLD, the events and the order of events in audio clips are known, without knowing the occurrence time of events. This paper proposes a connectionist temporal classification (CTC) based SED system that uses SLD instead of strongly labelled data, with a novel unsupervised clustering stage. Experiments on 41 classes of sound events show that the proposed two-stage method trained on SLD achieves performance comparable to the previous state-of-the-art SED system trained on strongly labelled data, and is far better than another state-of-the-art SED system trained on weakly labelled data, which indicates the effectiveness of the proposed two-stage method trained on SLD without any onset/offset time of sound events.
\end{abstract}
\section{Introduction}
\label{sec:intro}

Sound event detection (SED) aims to detect the class of acoustic events with the exact onset and offset time for the events. Classical applications of SED techniques include home monitoring and public security surveillance \cite{1}. 

Many SED methods typically rely on strongly labelled data, also known as frame level labelled data \cite{2, 3}. In strongly labelled data, each audio clip is labelled with both types of events in the audio clip and the onset/offset time of events. Based on strongly labelled data, the baseline system in \cite{4} feeds a block of frames into a convolutional neural network (CNN) to learn high-level features and recurrent neural network (RNN) to learn temporal information. One of the classical SED systems proposed by Adavanne and Virtanen \cite{5} (referred as \textsl{A\&V} system in the context) uses stacked convolutional and recurrent neural network as the main architecture and predicts labels at the frame level with a median filter used. Another type of SED method is based on weakly labelled data, also known as clip level labelled data \cite{6, 7}. In weakly labelled data, each audio clip is labelled with one or several tags of events in the audio clip without indicating the occurrence time and order information of events. Since no frame level information of sound events is provided in weakly labelled data, the whole audio clip is usually fed into models without dividing the clip into blocks \cite{8}. Using weakly labelled data, another classical SED system proposed by Xu et al. \cite{7} (referred as \textsl{XKWP} system in the context) uses intermediate variables of the model to infer the temporal locations of sound events to complete SED tasks. Due to the lack of frame level information, the SED algorithms with weakly labelled data cannot achieve a comparable performance with SED algorithms with strong labels.

Labelling strongly labelled data is time-consuming and labor expensive, so the size of strongly labelled dataset is often limited to a few minutes or hours \cite{6}. Though there are many weakly labelled datasets on the Internet, they are difficult to use in SED due to insufficient temporal information. Thus we proposed sequentially labelled data (SLD) \cite{9} inspired by the label sequence in speech recognition \cite{10}, where the sound events and the order of events are known in audio clips, without knowing the onset/offset time of events. With the help of connectionist temporal classification (CTC), SLD has been successfully applied for audio tagging \cite{9, 11}.

Previous work \cite{12} used CTC to solve SED problem, using time boundaries of events as labels. However, the time positions of events predicted by CTC are not close to the actual time boundaries when time boundaries labels sequences are used in the training phase. To solve this problem, \cite{12} uses the exact time boundaries of events as hints for the CTC model to find the actual time boundaries positions, which means strongly labelled data is used in \cite{12} rather than SLD.

In this paper, we extend our previous works \cite{9, 11} to do SED with SLD. The difficulty of solving SED problem based on SLD is to learn the location of time positions of different sound events in audio clips from the audio data without any onset/offset time of sound events.

This paper proposes a CTC-based SED system that uses SLD instead of strongly labelled data, with a novel unsupervised clustering stage. In Stage 1, sequential audio tagging is applied based on CTC using SLD to detect what events happen in audio clips and the order of events. In Stage 2, for each audio clip, frames of bottleneck features from the model trained in Stage 1 are clustered into either \textsl{background} cluster or \textsl{foreground} cluster using unsupervised clustering. The novelty of the proposed method is that event activity frames are obtained from the \textsl{foreground} cluster without using strongly labelled data. Then, combining sequential tags and events activity frames, the SED task can be completed.

This paper is organized as follows, Section 2 introduces the two-stage method in detail. Section 3 describes the dataset, baseline, experimental setup and analyzes the results. Section 4 gives conclusions.

\section{TWO-STAGE DETECTION METHOD}
\label{sec:format}

\begin{figure}[htb]
	\vspace{-0.2cm}  
	\setlength{\abovecaptionskip}{0.065cm}   
	\setlength{\belowcaptionskip}{-1.025cm}   
	\centerline{\includegraphics[width = 0.5  \textwidth]{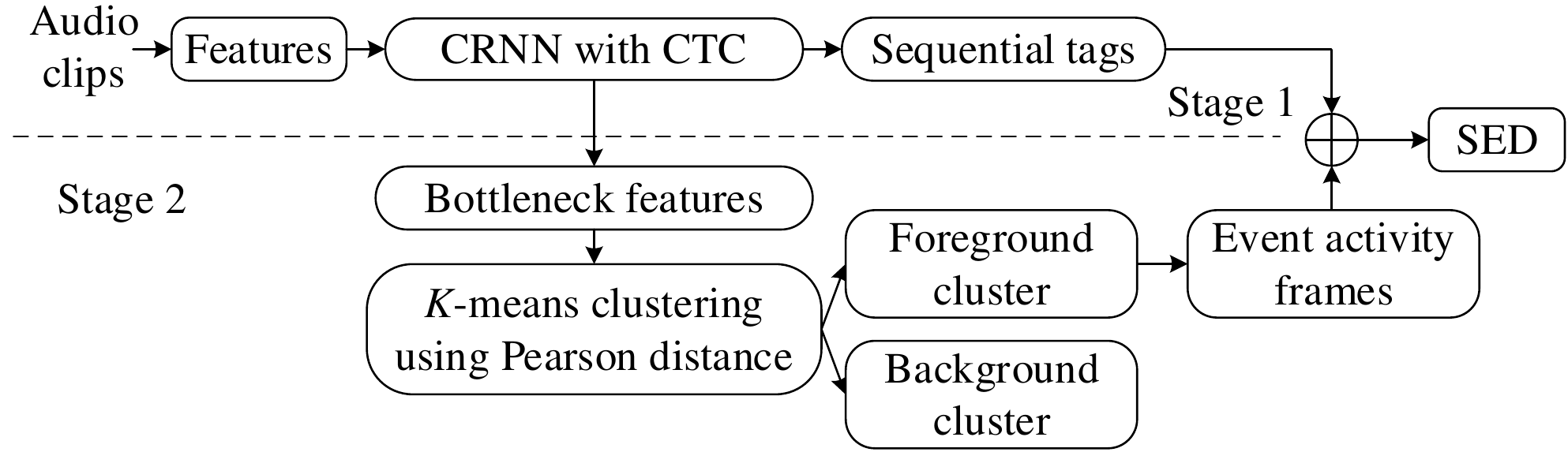}}
	\caption{Block diagram of the proposed method.}\medskip\label{fig1}
\end{figure}

A two-stage method is proposed for the SED problem. Stage 1 detects what events happened in audio clips and the order of events. In Stage 1, the events and the order of events in an audio clip are known, while their occurrence time remains unknown. Stage 2 detects the event activity frames, in which the event activity frames in an audio clip are known, without knowing what events are in the frame. Therefore, SED can be performed by combining the sequential tags in Stage 1 with the event activity frames in Stage 2, as shown in Fig. \ref{fig1}.
	
\subsection{Stage 1: Sequential audio tagging with SLD}
	
\label{ssec:stage1}
\begin{figure}[t]
	\setlength{\abovecaptionskip}{-0.0cm}   
	\setlength{\belowcaptionskip}{-0.8cm}   
	\centerline{\includegraphics[width = 0.5  \textwidth]{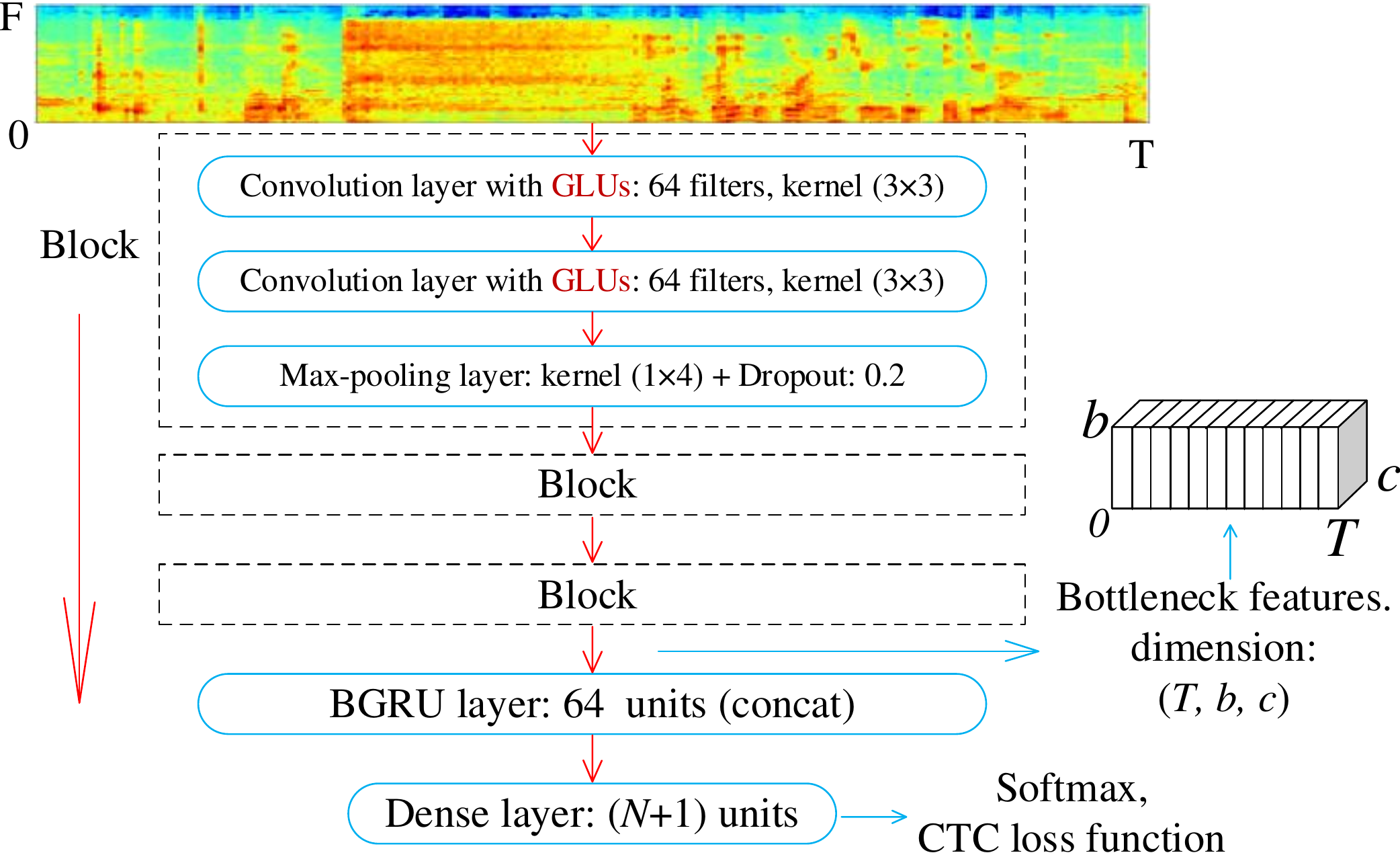}}
	\caption{CRNN-CTC of Stage 1. Note that the dimension of bottleneck features used in Stage 2 is (\textsl{T, b, c}), \textsl{T} is the number of input frames, \textsl{b} and \textsl{c} denote the frequency bins and the channel number of feature maps, respectively.
	}\medskip\label{fig2}
\end{figure}

To detect what events happen in audio clips and the order of events, sequential audio tagging using SLD is proposed in Stage 1. For the good performance of convolutional recurrent neural network (CRNN) in audio tagging \cite{8}, the CRNN is used as the basic classification model in Stage 1. For sequential audio tagging using SLD, CTC is used to keep the sequential information of events in model prediction. CTC \cite{13} redefines the loss function of a recurrent neural network (RNN) and allows the RNN to be trained for sequence-to-sequence tasks, without requiring any prior alignment between the input and target sequences \textsl{i.e.} the starting and ending time of sound events. As a result, it is sufficient to do audio tagging with SLD based on CRNN-CTC model.

Fig. \ref{fig2} shows the basic CRNN model trained with the CTC loss function. The waveforms of audio clips are converted to log mel spectrograms. Convolutional layers are applied to learn local shift-invariant patterns from features. To preserve the time resolution of the input, pooling is applied to the frequency axis only \cite{5}. Bidirectional gated recurrent units (BGRU) \cite{14} are adopted to capture the temporal context information. The final prediction layer has (\textsl{N}+1) units, where \textsl{N} is the number of sound event classes and the extra `1' indicates the blank label for CTC loss function \cite{13}.

To reduce the gradient vanishing problem in deep networks, gated linear units (GLUs) \cite{15} are proposed to replace the ReLU \cite{16} activation in the CRNN model. 
These provide a linear path for gradients propagation while keeping nonlinear capabilities through the sigmoid operation \cite{15}. 
GLUs can control the amount of information from a unit that flows to the next layer by sigmoid function. Given \textsl{W} and \textsl{V} are convolutional filters, \textsl{b} and \textsl{c} are biases, \textsl{X} denotes the input features in the first layer or the feature maps of the interval layers and $\sigma$ is sigmoid function, the GLUs can be defined as:
	\begin{equation}
		\setlength{\abovedisplayskip}{3pt}
		\setlength{\belowdisplayskip}{3pt}
		Y=(W \ast X + b)\odot \sigma (V \ast X +c)
	\end{equation}
where the symbol $\odot$ is the element-wise product and $\ast$ is the convolution operator. Another benefit of using GLUs is that network can learn to attend to sound events and ignore the unrelated sounds. If the value of sigmoid function is close to 1, then the corresponding Time-Frequency unit is attended.

\subsection{Stage 2: Unsupervised clustering}
\label{ssec:stage2}

To obtain the event activity frames of each audio clip, the bottleneck features of each audio clip from Stage 1 are clustered to two clusters: a \textsl{background} cluster and a \textsl{foreground} cluster. The \textsl{background} means the background acoustic scene of the audio clip, and \textsl{foreground} cluster is regarded as a cluster of multiple sound events in the audio clip.

For each audio clip, suppose \textsl{F} is the bottleneck feature output from the CRNN trained in Stage 1, where \{$F_1,...,F_T$\} are the frames of \textsl{F}. For each frame $F_i$, if it contains target sound events, it will be regarded as a \textsl{foreground} frame; otherwise, it will be regarded as a \textsl{background} frame. Since there are no frame level labels in SLD, an unsupervised \textsl{K}-means clustering algorithm \cite{17} is used to obtain the \textsl{background} cluster and \textsl{foreground} cluster from the bottleneck features of each audio clip, which means \textsl{K} equals 2 in \textsl{K}-means. Most \textsl{K}-means algorithms use Euclidean distance, also known as $L_2$ norm. However, Euclidean distance is sensitive to outliers \cite{18}. To better measure the distance among frames, Pearson distance \cite{19} is used as an alternative distance function in clustering. The performance of clustering based on Euclidean distance and Pearson distance will be investigated in experiments. The Pearson distance between two frames can be defined as:
\begin{equation}
	\setlength{\abovedisplayskip}{3pt}
	\setlength{\belowdisplayskip}{3pt}
	d_{\textit{pearson}}(F_m,F_n)=1-\rho(F_m,F_n)
\end{equation}
where $\rho$ is the Pearson Correlation Coefficient (PCC) \cite{20} of \textsl{m}-th frame and \textsl{n}-th frame. Considering that the PCC falls between [$-$1, 1], Pearson distance lies in [0, 2].

After unsupervised clustering, there are two clusters of frames. In real life, acoustic scene of audio recording is unlikely to change too suddenly or frequently, but different sound events in an audio clip may vary greatly. Therefore, the distance among frames of \textsl{background} cluster in the audio clip should be smaller, the \textsl{background} cluster should be more compact than the \textsl{foreground} cluster. To evaluate which cluster is more compact, the average Euclidean distance is calculated between each frame of the cluster and the cluster centroid, and the smaller one is regarded as the \textsl{background} cluster. Given frames \{$F_1,...,F_n$\} in cluster \textsl{C}, the average Euclidean distance is calculated by:
\begin{equation}
	\setlength{\abovedisplayskip}{3pt}
	\setlength{\belowdisplayskip}{3pt}
	d_{\textit{avg}}=\sum\nolimits_{i=1}^nd(F_i,p)/n
\end{equation}
where \textsl{d} is the Euclidean distance and \textsl{p} is the cluster centroid.

By comparing the average Euclidean distance of two clusters in an audio clip, the cluster that has smaller average Euclidean distance is regarded as \textsl{background} cluster, and another is \textsl{foreground} cluster. From the \textsl{foreground} cluster, the frames of acoustic events are detected. Note that the event activity frames are obtained from \textsl{foreground} cluster, we only know that there are sound events in the event activity frames, without knowing what it is in each frame.

\subsection{Combining sequential tags and event activity frames}
\label{ssec:combining}

Given \{$F_1,...,F_N$\} are the acoustic event frames detected by \textsl{foreground} cluster in Stage 2, \{$S_1,...,S_K$\} are used to present the event spikes sequence \textsl{S} in an audio clip whose corresponding time positions are presented as \{$Ts_1,...,Ts_K$\}, respectively. The spikes $S$ usually occur near the maximum posterior probability of the events, which is located within the period of acoustic event occurrence \cite{13}. Therefore, the distance between each spike time position $Ts_j$ and each activity frame index $F_i$ is calculated, and each activity frame is assigned to the nearest event spike, which means the label $S_j$ of each activity frame $F_i$ can be defined as:
\begin{equation}
	\setlength{\abovedisplayskip}{2.5pt}
	\setlength{\belowdisplayskip}{2.5pt}
	S_j=\mathop{\arg\min}\limits_{j}\mid F_i-Ts_j\mid,j=1,...,K;i=1,...,N
\end{equation}

At this point, both the event activity frames of audio clips and the classes of events in each frame are identified, which marks the completion of SED tasks.

\section{EXPERIMENTS AND RESULTS}
\label{sec:EXPERIMENTS}

\vspace{-0.075cm}
\subsection{Dataset, Baseline and Experiments Setup}
\label{ssec:Dataset}

Previous work \cite{12} tested the CTC method in SED on 17 types of sound events. To evaluate our proposed two-stage method on more types of sound events, the DCASE 2018 Task 2 dataset \cite{21} is used in this paper. Task 2 contains 41 kinds of sound events from \textsl{Freesound}, and is larger than the datasets in other DCASE tasks \cite{4, 22, 23}. These sound events are remixed with acoustic scenes into 10-second audio clips, where each audio clip contains 2 to 4 sound events mixed with TUT Urban Acoustic Scenes recordings \cite{24}. The source code for synthesizing data can be found at \cite{25}, and the signal-to-noise ratio (\textsl{SNR}) of synthesizer is 0 dB. The mixed target events are non-overlapped. Target events and non-target events in acoustic scenes overlap.

For baselines, the \textsl{A\&V} system \cite{5} is trained on strongly labelled data and \textsl{XKWP} system \cite{7} is trained on weakly labelled data. The proposed method is trained on SLD. SLD is derived from the strongly labelled data following \cite{9}, using the sequence of events as labels. For polyphonic audio clips such as \textsl{a dog barks while the bell rings}, although the two events overlap, the label of this audio clip can still be (\textsl{ringing, bark}). These methods are evaluated using the synthetic large-scale dataset totalling 33.4 hours.
	
In the training phase, log mel-band energy is extracted using STFT with a Hamming window of 64 ms length, which has a sufficient time and frequency resolution of spectrum. The overlap of 50\% between the window is used to smooth the spectrogram. Then 64 mel filter banks are applied \cite{25}. Dropout and Early-stopping are used to prevent over-fitting. The Adam \cite{26} optimizer is used with a learning rate of 0.001. Four-fold cross-validation is used for model selection.

\vspace{-0.075cm}
\subsection{Results and Analysis}
\label{ssec:Results}

\begin{figure}[tb]
	\setlength{\abovecaptionskip}{0.10cm}   
	\setlength{\belowcaptionskip}{-0.8cm}   
	\centerline{\includegraphics[width = 0.5  \textwidth]{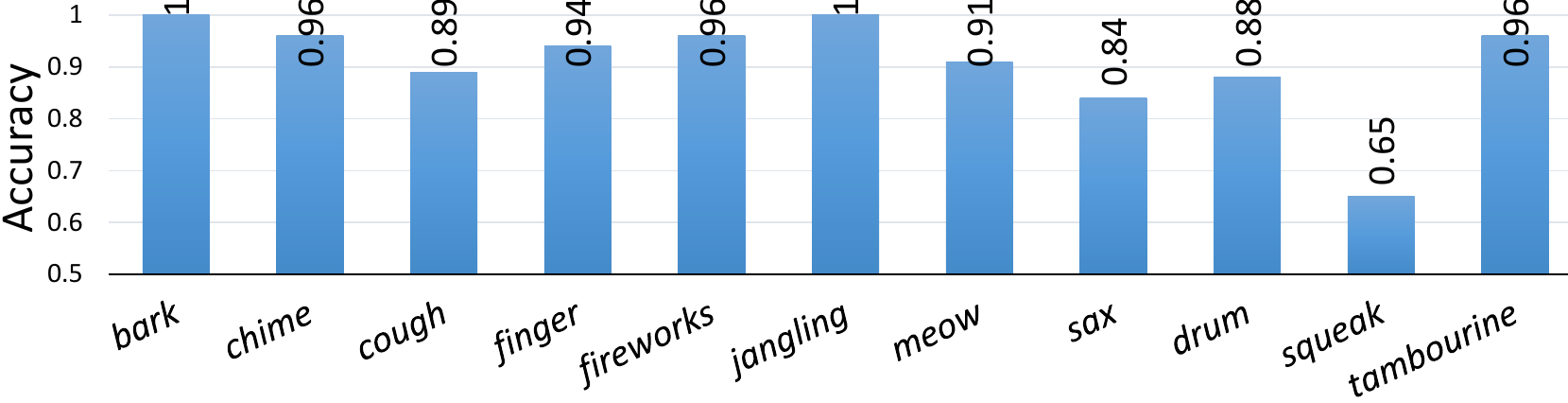}}
	\caption{Audio tagging class-wise accuracy.}\medskip\label{fig3}
\end{figure}

The metrics of \textsl{Precision} (\textsl{P}), \textsl{Recall} (\textsl{R}), \textsl{F-score} and \textsl{Error rate} (\textsl{ER}) \cite{27} are based on segments of 1 second length. Higher \textsl{P}, \textsl{R}, \textsl{F-score} and lower \textsl{ER} indicate better performance.
The audio tagging results can show the performance of CRNN model in Stage 1, and also indirectly reflect the quality of the bottleneck features learned by the CRNN. Only if the model learns better high-level acoustic features of audio clips, will the model better detect sound events. To provide an intuition on how well the approach performs on individual classes, the audio tagging accuracy is shown in Fig. \ref{fig3}. Due to the limitation of space, 11 classes of events are randomly selected in Fig. \ref{fig3}, for the detailed results of all 41 classes of events, please see here\protect\footnote{https://github.com/moses1994/SED\_based\_on\_SLD}.

In Fig. \ref{fig3}, events such as `bark' and `jangling' have 100\% classification accuracy but some classes like `squeak' have poor performance. A reason may be that `squeak' sound varies, depending on the objects caused them. In detail, we found that `squeak' and `bass' are often confused with many other classes, and `fart' is often confused with `sax'. Another interesting pair is `computer keyboard' and `drum', it seems reasonable to confuse them as they sometimes sound to be similar. Confusion matrix for all 41 classes of events is available online$^1$. Averaged \textsl{P}, \textsl{R} and \textsl{F-score} of audio tagging is \textsl{95.03\%}, \textsl{88.81\%} and \textsl{91.81\%}, respectively. We see that the CRNN-CTC trained on SLD performs well in audio tagging.	

To evaluate the classification accuracy of the \textsl{background} cluster and \textsl{foreground} cluster based on Euclidean distance and Pearson distance, the classification precision at the cluster level is calculated. According to the frame level ground-truth test data labels, the classification precision of unsupervised clustering based on Euclidean distance and Pearson distance is \textsl{70.33\%} and \textsl{88.62\%}, respectively. Consequently, the clustering results based on Pearson distance are used for SED.

For SED, the proposed method (referred as \textsl{HKLP}) is compared with \textsl{A\&V} and \textsl{XKWP}. As shown in Table 1, the proposed method trained on SLD achieves performance close to \textsl{A\&V} and is far better than \textsl{XKWP}. The standard deviation (\textsl{SD}) of \textsl{ER} for all 41 classes of \textsl{A\&V}, \textsl{XKWP} and the proposed method is \textsl{0.24}, \textsl{0.22} and \textsl{0.11}, respectively. Although the \textsl{ER} of the proposed method is \textsl{10\%} worse than the \textsl{A\&V} trained on strongly labelled data, the proposed method reduced the standard deviation of \textsl{ER} by half. The proposed method is more stable for 41 different class events.

The SED class-wise \textsl{ER} are shown in Fig. \ref{fig4}, for all 41 classes of events are available here$^1$. For some classes, the performance of our method (green) is similar to \textsl{A\&V} (grey). For others like `bark', `cough' and `finger', our method is better. For overall evaluation in Table 1, lower \textsl{ER}, \textsl{deletion (D) rate}, \textsl{insertion (I) rate} and \textsl{substitution (S) rate} \cite{27} indicate better performance. \textsl{D rate} means the ratio of events that are not recognized to the total events in the ground-truth. In Table 1, the \textsl{D rate} is the main error rate in the overall \textsl{ER} of the proposed method. To study the reason for this phenomenon, the bottleneck features and event activity frames in an audio clip are shown in Fig. \ref{fig5}. Note that the \textsl{SNR} of synthesizer is 0 dB, so there are many other non-target events in the spectrogram. The event spike sequence is correct, but the event activity frames do not match well with the ground-truth red lines. Some frames where events occur are not recognized by the unsupervised clustering in Stage 2. This may be the reason for \textsl{D rate} is the main error rate in the \textsl{ER} of the proposed method.

\begin{table}[H]\footnotesize
	\setlength{\abovecaptionskip}{0cm}   
	\setlength{\belowcaptionskip}{0cm}   
	\renewcommand\tabcolsep{0.5pt} 
	\centering
	\caption{Evaluation of SED among methods for 41 classes.}
	\begin{tabular}{p{1cm}<{\centering}p{0.7cm}<{\centering}p{0.9cm}<{\centering}p{0.9cm}<{\centering}p{0.9cm}<{\centering}p{1.1cm}<{\centering}p{1.2cm}<{\centering}p{1.4cm}<{\centering}} 
		\toprule[1pt]
		\specialrule{0em}{0.1pt}{0.1pt}
		& \textsl{ER} & \textsl{D rate} & \textsl{I rate} & \textsl{S rate} & \textsl{F-score} &\textsl{SD of ER}&\textsl{Label type}\\
		\midrule[1pt]  
		\specialrule{0em}{0.1pt}{0.1pt}
		\textsl{\textsl{HKLP}}&0.46 & \textbf{0.303} & 0.101 & 0.058 & 70.98\% & \textbf{0.11} &\textsl{sequential}\\
		\specialrule{0em}{0.2pt}{0.2pt}
		\textsl{A\&V} & \textbf{0.40} & 0.347 & \textbf{0.030} & \textbf{0.021} & \textbf{75.05\%} & 0.24 & \textsl{strong}\\
		\specialrule{0em}{0.2pt}{0.2pt}
		\textsl{XKWP} & 0.94 & 0.791 & 0.102 & 0.044 & 25.02\% & 0.22 &\textsl{weak}\\
		\specialrule{0em}{0.2pt}{0.2pt}
		\bottomrule[1pt]
		\specialrule{0em}{0pt}{0pt}
	\end{tabular}
	\label{tab:table}
\end{table}

\begin{figure}[t]
	\setlength{\abovecaptionskip}{0.2cm}   
	\setlength{\belowcaptionskip}{-0.5cm}   
	\centerline{\includegraphics[width = 0.5  \textwidth]{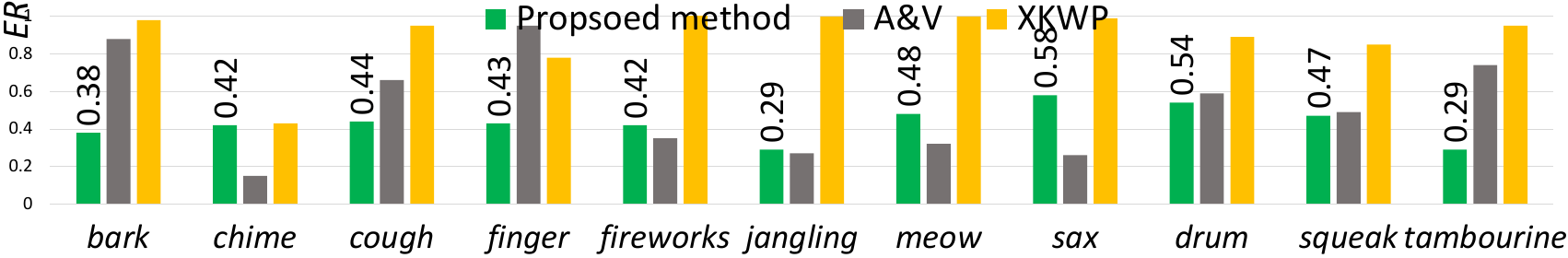}}
	\caption{The results of \textsl{ER} in frame level, the specific value of the proposed method is shown. Lower \textsl{ER} indicates better.}\medskip\label{fig4}
\end{figure}
\setlength{\parindent}{0.0em}



\vspace{-1.05em}
\begin{figure}[htb]
    \setlength{\abovecaptionskip}{0.9em}   
	\setlength{\belowcaptionskip}{-0.9em}   
	\centerline{\includegraphics[width = 0.5  \textwidth]{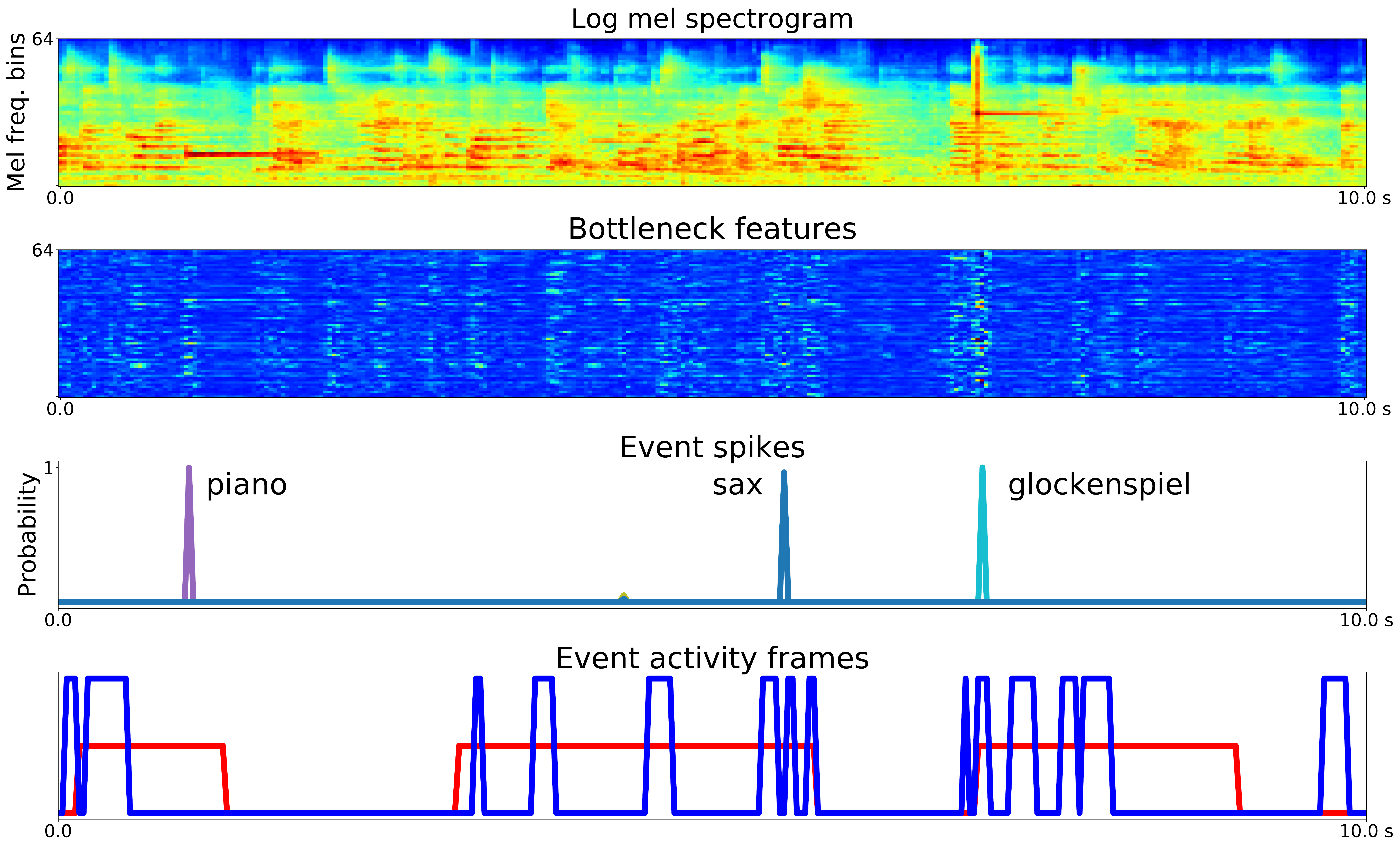}}
	\caption{From top to bottom: log mel spectrogram, bottleneck features, event spikes sequence and event activity frames. In the bottom subgraph, the red lines and blue lines denote the ground-truth event activity frames and the proposed event activity frames in Stage 2, respectively.}\medskip\label{fig5}
\end{figure}
	
\section{CONCLUSION}
\label{sec:CONCLUSION}

This paper proposed a CTC-based SED system that uses SLD instead of strongly labelled data, with a novel unsupervised clustering stage. Experimental results show the performance of the proposed method using SLD is comparable to the previous \textsl{A\&V} system using strongly labelled data, and is far better than the \textsl{XKWP} system using weakly labelled data, indicating the effectiveness of the proposed method.

\setlength{\parindent}{2.0em}
For the proposed method, the main error rate in \textsl{ER} is the \textsl{D rate} because the proposed event activity frames based on unsupervised clustering in Stage 2 are still inaccurate, and time location of sound events in audio clips is difficult in SED tasks. The future work will focus on improving the accuracy of proposed event onset/offset time in audio clips.	

\section{ACKNOWLEDGEMENTS}
\label{sec:ACKNOWLEDGEMENTS}
This research was partly supported by EPSRC grant EP/N014111/1 ``Making Sense of Sounds" and a Research Scholarship from the China Scholarship Council (CSC) No. 201406150082.

\vfill\pagebreak


\bibliographystyle{IEEEbib}
\bibliography{icassp2019}

\end{document}